\documentclass[prd,aps,lengthcheck,twocolumn,nofootinbib,notitlepage,floatfix,superscriptaddress]{revtex4-2}
\usepackage{graphics,graphicx}
\usepackage{amsmath, amssymb}
\usepackage{multirow}
\usepackage{color}
\usepackage{verbatim}
\usepackage{hyperref}
\usepackage[normalem]{ulem}  
\usepackage{ulem}
\usepackage{color}
\usepackage{cancel}
\usepackage{mathtools}

\renewcommand\sout{\bgroup\color{blue} \ULdepth=-.5ex \ULset}
\def\slashchar#1{\setbox0=\hbox{$#1$}  
\dimen0=\wd0     
\setbox1=\hbox{/} \dimen1=\wd1  
\ifdim\dimen0>\dimen1   
\rlap{\hbox to \dimen0{\hfil/\hfil}} 
#1     
\else     
\rlap{\hbox to \dimen1{\hfil$#1$\hfil}} 
/      
\fi}

\newcommand{\eps}{\epsilon}
\newcommand{\dd}{\mathrm{d}}
\newcommand{\pp}{\partial}

\begin{document}

\title{Challenges in locating the QCD critical point via constant entropy density contours}

\date{\today}

\author{Micha\l{} Marczenko}
\email{michal.marczenko@uwr.edu.pl}
\affiliation{Incubator of Scientific Excellence - Centre for Simulations of Superdense Fluids, University of Wroc\l{}aw, plac Maksa Borna 9, PL-50204 Wroc\l{}aw, Poland}
\author{Micha\l{} Szyma\'nski}
\email{michal.szymanski@uwr.edu.pl}
\affiliation{Institute of Theoretical Physics, University of Wroc\l{}aw, plac Maksa Borna 9, PL-50204 Wroc\l{}aw, Poland}

\author{Gy\H{o}z\H{o} Kov\'acs}
\email{gyozo.kovacs@uwr.edu.pl}
\affiliation{Institute of Theoretical Physics, University of Wroc\l{}aw, plac Maksa Borna 9, PL-50204 Wroc\l{}aw, Poland}
\affiliation{Institute for Particle and Nuclear Physics, HUN-REN Wigner Research Centre for Physics, 1121 Budapest, Konkoly–Thege Miklós út 29-33, Hungary}

\begin{abstract}
A new method was proposed recently to investigate the location of the putative critical point of strongly interacting matter, governed by quantum chromodynamics. By approximating contours of constant entropy density at finite baryon chemical potential, the conditions for the existence of a critical point are solved. In this work, we analyze this method in the hadron resonance gas and Nambu--Jona-Lasinio models. We demonstrate that the prediction of the critical point in the HRG model is solely due to mesonic and baryonic degrees of freedom, and thus is not necessarily a signal of a critical point. We argue that such an expansion leads to a physically meaningful prediction only when applied near the critical point.
\end{abstract}
\maketitle

\section{Introduction}
\label{sec:intro}

One of the key tasks addressed in the context of high-energy physics is the determination of the phase diagram of Quantum Chromodynamics (QCD) at finite temperature and baryon chemical potential. At low baryon chemical potential and high temperature, the transition from hadronic to quark-gluon plasma phase is smooth and is connected with the restoration of chiral symmetry~\cite{Bazavov:2014pvz, Borsanyi:2018grb, Bazavov:2017dus, Bazavov:2020bjn, Bazavov:2020bjn, Aoki:2006we}. On the other hand, based on the effective model calculations, one expects a first-order phase transition at large baryon chemical potential and low temperature~\cite{Bowman:2008kc, Ferroni:2010ct, Buballa:2003qv, Sasaki:2007qh}. This would imply the presence of a putative critical point (CP) on the QCD phase diagram. Throughout recent years experimental attempts have been made to locate it on the QCD phase diagram. Despite enormous experimental effort within the beam energy scan (BES) programs at the Relativistic Heavy Ion Collider (RHIC) at BNL~\cite{STAR2010} and the Super Proton Synchrotron (SPS) at CERN~\cite{Mackowiak-Pawlowska:2020glz}, this pressing issue remains unresolved (for a recent review see~\cite{Bzdak:2019pkr}).

A new method for locating the QCD CP was recently proposed~\cite{Shah:2024img}. It is based on the fact that thermodynamic quantities, such as entropy, are multi-valued between the spinodals of the first-order phase transition. This leads to the crossing of lines of different constant entropies in the first-order region. Thus, constant-entropy contours at finite temperature and baryon chemical potential can be used to locate the onset of a first-order phase transition and, therefore, the location of a critical endpoint. In~\cite{Shah:2024img}, the constant-entropy contours were approximated via the Taylor series at vanishing baryon chemical potential to predict the existence of a CP. The constant-entropy-contour scheme was recently used in LQCD simulations to exclude the location of CP at $\mu_B < 450~$MeV at $2\sigma$ confidence level~\cite{Borsanyi:2025dyp}.

In this work, we apply the constant-entropy expansion scheme to the hadron resonance gas (HRG) and Nambu--Jona-Lasinio (NJL) models to test its predictive power for the CP. Both models are extensively studied at finite temperature and baryon chemical potential. The HRG model has been successful in describing LQCD data on the EoS and fluctuations of conserved charges at vanishing baryon chemical potential~\cite{HotQCD:2014kol, Bollweg:2021vqf} as well as the hadron yields from heavy-ion collisions~\cite{Andronic:2017pug}. The NJL model describes the effective interactions of quarks preserving the chiral symmetry of the massless QCD Lagrangian. We demonstrate that the constant-entropy expansion scheme leads to the unphysical prediction of the existence of CP in the HRG model, which does not implement any criticality. In addition, we directly compare the results obtained in the full NJL model (which may or may not predict a CP, depending on its details), with the prediction obtained through the constant-entropy expansion scheme.

This paper is organized as follows. In Sec.~\ref{sec:expansion}, we introduce the constant-entropy expansion scheme. In the following two sections, we discuss the results obtained within the HRG and NJL models. The last section concludes our findings.

\section{Taylor expansion of the constant-entropy contours} \label{sec:expansion}

\begin{figure}
    \centering
    \includegraphics[width=1\linewidth]{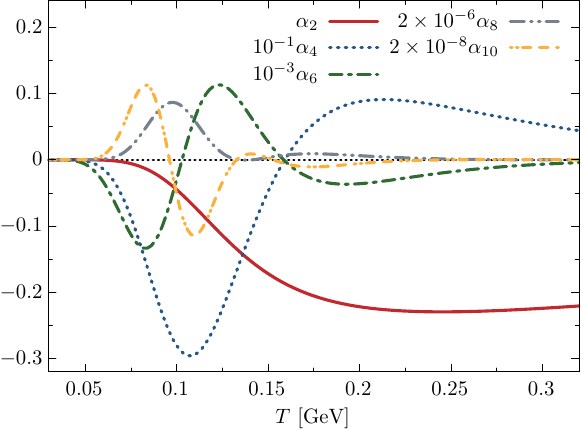}
    \caption{Temperature dependence of the expansion coefficients, $\alpha_2$, $\alpha_4$, $\alpha_6$, $\alpha_8$, and $\alpha_{10}$. Note that the coefficients are rescaled for better presentation.}
    \label{fig:alpha_orders}
\end{figure}

The method proposed in~\cite{Shah:2024img} considers the contours of constant entropy density, $s$, at finite $T$ and $\mu_B$, i.e.,
\begin{eqnarray}
    s\left(T_s\left(T, \mu_B\right), \mu_B\right) = s\left(T, \mu_B=0\right)
\end{eqnarray}
The temperature contour $T_s\left(T, \mu_B\right)$ is chosen such that the entropy is fixed at arbitrary $\mu_B$. The profile $T_s(T, \mu_B)$ can be expanded into Taylor series around $\mu_B=0$,
\begin{equation}\label{eq:taylor}
    T_s(T, \mu_B) = T + \sum_{k=1}^\infty \frac{\alpha_{k}}{k!} \mu_B^{k} \textrm,
\end{equation}
where
\begin{equation}\label{eq:alpha}
    \alpha_k = \frac{\dd^k T}{\dd \mu_B^k} \Bigg|_s \textrm.
\end{equation}
It is understood that the coefficients $\alpha_k$ depend on the temperature $T$, i.e., $\alpha_k = \alpha_k(T)$. Since odd coefficients $\alpha_{2k+1}$ vanish at $\mu_B=0$, the lowest-order non-vanishing coefficient is $\alpha_2$, which reads
\begin{eqnarray} \label{eq:alpha2}
    \alpha_2 = - \frac{\partial^2 s}{ \partial \mu_B^2}\Bigg|_T \; \Bigg/ \;\frac{\partial s } {\partial T}\Bigg|_{\mu_B} \textrm,
\end{eqnarray}
where 
\begin{eqnarray}
    s = \frac{\partial P}{\partial T}\Bigg|_{\mu_B}
\end{eqnarray}
is the entropy density. The higher-order coefficients are more complicated functions of entropy and its derivatives with respect to the temperature and/or baryon chemical potential. In general, the following recursive relation holds for $k>0$:
\begin{equation}
    \alpha_{k+1} =  \frac{\dd^{k+1} T}{\dd \mu_B^{k+1}} \Bigg|_s = \frac{\dd \alpha_k}{\dd \mu_B}  \Bigg|_s = \frac{\pp \alpha_k}{\pp \mu_B} \Bigg|_T + \alpha_1 \frac{\pp \alpha_k}{\pp T} \Bigg|_{\mu_B} \textrm,
\end{equation}
from which $\alpha_k$ for $k>2$ can be systematically derived starting from Eq.~\eqref{eq:alpha2}. 

In principle, the critical point can be located through the equations
\begin{align}\label{eq:CP_criterion}
    \frac{\partial T_s}{\partial s}\Bigg|_{\mu_B} = 0 \textrm, \quad
\frac{\partial^2 T_s}{\partial s^2}\Bigg|_{\mu_B} = 0 \textrm.
\end{align}

Truncating the Taylor series at the lowest-order ($k=2$), the criteria can be written with the temperature derivatives of $\alpha_2$,
\begin{equation}\label{eq:CP_criterion_alpha}
\begin{split}
    2 + \frac{\dd \alpha_2}{\dd T} \mu_B^2 = 0 \textrm,\quad
    \frac{\dd^2 \alpha_2}{\dd T^2} = 0 \textrm.
\end{split}
\end{equation}
Solving these equations yields a prediction for the location of the CP in terms of the critical temperature $T_c$ and baryon chemical potential $\mu_{B,c}$~\cite{Shah:2024img},
\begin{eqnarray}
    \mu_{B,c}&=&\sqrt{-\frac{2}{\alpha_2'(T_{0,c})}}\nonumber\\
    T_{c}&=&T_{0,c}+\alpha_2(T_{0,c})\frac{\mu_{B,c}^2}{2}\,,
\end{eqnarray}
where $T_{0,c}$ is obtained from the second of Eqs.~\eqref{eq:CP_criterion_alpha}. We note that the criteria in Eq.~\eqref{eq:CP_criterion_alpha} predict the existence of the CP at the real baryon chemical potential for $\dd \alpha_2 / \dd T < 0$, that is, when $\alpha_2$ features an inflection point with a negative slope. 

\begin{figure}
    \centering
    \includegraphics[width=1\linewidth]{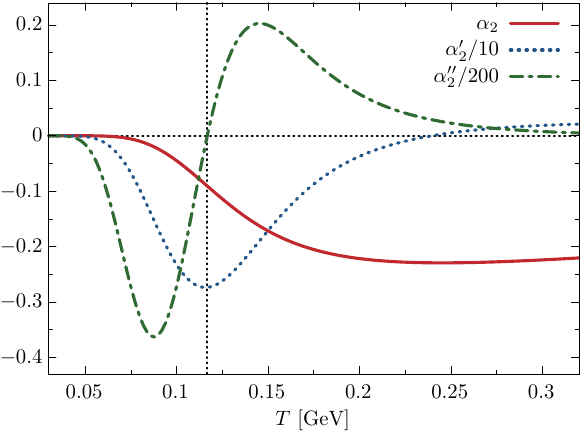}
    \caption{The second order expansion coefficient $\alpha_2$, its first ($\alpha_2'$) and second ($\alpha_2''$) derivatives w.r.t temperature. The vertical, dotted line marks the temperature at which the second derivative vanishes, i.e., $\alpha_2''=0$. The first and second derivatives are rescaled for better presentation.}
    \label{fig:alpha_der}
\end{figure}

\section{Hadron Resonance Gas Model}\label{sec:hrg}

In the phenomenological description of hadronic matter, one needs to identify the relevant degrees of freedom and the interactions among them. In the confined phase of QCD, the medium is composed of hadrons and their resonances. In its simplest version, the hadron resonance gas (HRG) model assumes that the constituents of the medium are independent and point-like~\cite{Braun-Munzinger:2003pwq}. This effectively neglects their widths and interactions. Consequently, the pressure in the HRG model is approximated by the sum over partial pressures of hadrons and their resonances, treated as noninteracting particles,
\begin{equation}\label{eq:hrg_pressure}
    P=\sum\limits_{i} P_i \rm,
\end{equation}
where $i$ goes through all strange and non-strange hadrons and their resonances listed in the Particle Data Group summary tables~\cite{ParticleDataGroup:2022pth}\footnote{In this work, we include established mesons and baryons with three- and four-star rating.}. We note that the thermodynamic pressure $P$ contains all the relevant information about the medium through the mass and the quantum numbers of hadrons. Thus, it allows for the study of different thermodynamic observables, including particle numbers and fluctuations of conserved charges.

\begin{figure}
    \centering
    \includegraphics[width=1\linewidth]{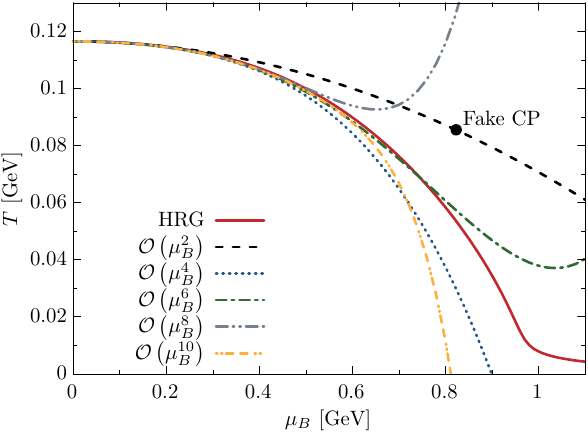}
    \caption{Constant-entropy contour obtained in the HRG model starting at $\mu_B=0$ and $T$ such that $\alpha_2'' = 0$, which leads to the prediction of a fake CP, shown as a black circle on the contour obtained with the $\mathcal{O}\left(\mu_B^2\right)$ truncation of the Taylor series defined in Eq.~\eqref{eq:taylor} (see text for details). Shown are also contours obtained with higher-order truncations up to $\mathcal{O}\left(\mu_B^{10}\right)$.}
    \label{fig:contour}
\end{figure}

The partial pressures in Eq.~\eqref{eq:hrg_pressure} are given as
\begin{equation}\label{eq:p_i}
    P_{i} = \mp \gamma_i T \int \frac{\dd^3p}{(2\pi)^3}\; \ln\left(1\mp f_i\right)\textrm,
\end{equation}
where $T$ is temperature, $\gamma_i$ is the spin degeneracy factor, and 
\begin{equation}\label{eq:distr}
    f_i = \left(e^{(\eps_i-\mu_i)/T} \pm 1\right)^{-1}
\end{equation}
is the distribution function, $\eps_i=\sqrt{\boldsymbol{p}^2+m_i^2}$ is the dispersion relation and \mbox{$\mu_i=B_i \mu_B+Q_i\mu_Q+S_i\mu_S$} is the chemical potential of the particle. In this work, we set $\mu_Q = \mu_S = 0$. The upper (lower) signs in Eqs.~\eqref{eq:p_i} and \eqref{eq:distr} refer to fermions (bosons).

We use the HRG model to calculate the Taylor expansion coefficients $\alpha_k$, defined in Eq.~\eqref{eq:alpha}. In Fig.~\ref{fig:alpha_orders}, we show the non-vanishing coefficients up to $k=10$ at $\mu_B=0$. Higher-order coefficients become more complex. For convenience, we denote the expansion truncated at order $k$ by superscript $(k)$. We are primarily interested in the expansion of the HRG model truncated at $k=2$. 

To locate the putative CP, one needs to solve Eqs.~\eqref{eq:CP_criterion_alpha}. The first and second temperature derivatives of $\alpha_2$ are shown in Fig.~\ref{fig:alpha_der}. We find that $\dd^2 \alpha_2 / \dd T^2 = 0$ at $T_0^c=0.117~$GeV. This yields $\left(T_c^{(2)}, \mu_{B,c}^{(2)}\right) = (0.086, 0.823)~\rm GeV$ for the location of the CP. We stress that the HRG model does not predict any CP because there is no criticality built into this model. Therefore, the second-order expansion leads to a nonphysical approximation of the full HRG model. This is shown in Fig.~\ref{fig:contour}, where we show the constant-entropy contour that leads to the fake CP. This is in contrast with the full HRG contour, which deviates from the truncated one as the baryon chemical potential increases. At $\mu_{B,c}^{(2)} = 0.823~$GeV, the truncated contour lies at a temperature higher than the one obtained in the full HRG by $60\%$. The deviation of the contours starts around $\mu_B\sim300-400$ MeV for the quadratic expansion. This is compatible with the results of Ref.~\cite{Borsanyi:2025dyp}, although they found that a $T_s(\mu_B)\propto\mu_B^2$ description is sufficient over a slightly broader range of (imaginary) $\mu_B$. 
Interestingly, we also find the contour from a quadratic truncation to be valid until larger values of $\mu_B$ for larger $T_0$, where the lattice calculation is performed.

\begin{figure}
    \centering
    \includegraphics[width=1\linewidth]{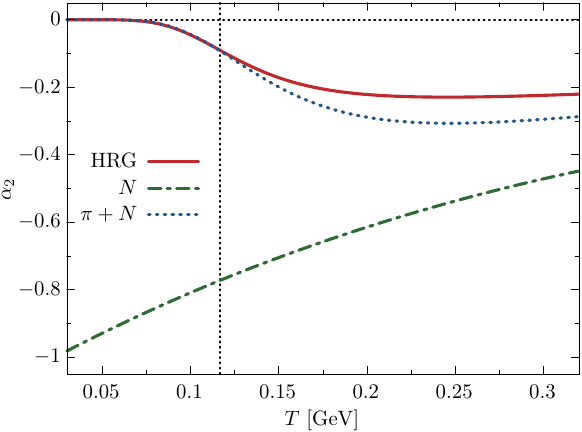}
    \caption{The coefficient $\alpha_2$ for the HRG model, as well as a gas of nucleons and a gas of pions and nucleons. The vertical, dotted line marks the temperature at which the second derivative vanishes, i.e., $\alpha_2''=0$ in the full HRG model.}
    \label{fig:alpha_example}
\end{figure}

We also considered the expansion truncated at the fourth order of the expansion, i.e., $k=4$. This model leads to $\left(T_c^{(4)}, \mu_{B,c}^{(4)}\right) = (0.066, 0.843)~\rm GeV$. Thus, even the fourth-order expansion leads to a nonphysical prediction of the location of the CP. Accordingly, the contour obtained from fourth (or even higher) order expansion deviates from the true contour at only slightly higher chemical potential compared to the quadratic truncation, as shown in Fig.~\ref{fig:contour}. This possibly indicates slow convergence of the expansion.

To better understand why the second-order expansion leads to an unphysical CP, we consider a single-component ideal gas. In this case, one may show that
\begin{eqnarray}\label{eq:limit}
    \alpha_2 \xrightarrow{T \rightarrow 0} -\frac{B^2}{M} \textrm,
\end{eqnarray}
where $M$ is the particle's mass and $B$ is its baryon number. Thus, $\alpha_2 \rightarrow 0$ for mesons and $\alpha_2 \rightarrow - M^{-1}$ for baryons as the temperature approaches zero. The limit in Eq.~\eqref{eq:limit} can be straightforwardly generalized to a multi-component gas of ideal particles, in which case the limit is determined by the lightest hadron.

The behavior of $\alpha_2$ in the system of nucleons is shown in Fig.~\ref{fig:alpha_example}. It is an increasing function of temperature. Therefore, the solution of Eqs.~\eqref{eq:CP_criterion_alpha} does not lead to any prediction of CP at finite baryon chemical potential. On the other hand, for a system of pions and nucleons, $\alpha_2$ starts from zero, decreases, and develops a minimum at $T > 0.2~$GeV. Consequently, it features an inflection point, and the solution of Eqs.~\eqref{eq:CP_criterion_alpha} leads to a prediction of CP at $\left(T_c^{(\pi N)}, \mu_{B,c}^{(\pi N)}\right) = (0.086, 0.816)~\rm GeV$. We note that the values are very close to those obtained in the truncated HRG model with the full resonance spectrum. This is so because at $T\lesssim 0.12~$GeV the thermodynamic pressure is predominantly due to pions and nucleons. This reassures that the inflection point of $\alpha_2$ that leads to the prediction of an unphysical CP in the full HRG model is obtained primarily due to the presence of pions and nucleons in the particle spectrum and should not depend on the list of resonances used in the calculations. We have verified this for the list of resonances that includes the prediction from the quark model~\cite{Loring:2001ky, Ebert:2009ub}.

\begin{figure}
    \centering
    \includegraphics[width=1\linewidth]{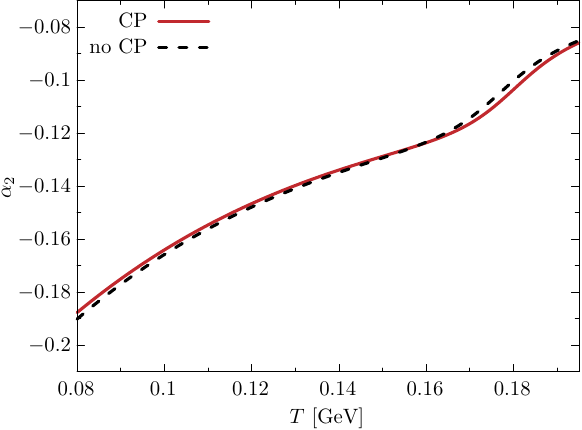}
    \caption{The expansion parameters $\alpha_2$ in two different schemes, one with (solid red line) and one without a CP (dashed black line).}
    \label{Fig:alpha_NJL_schemes}
\end{figure}

\section{Nambu--Jona-Lasinio model} \label{sec:NJL}

The NJL model Lagrangian for two quark flavors with a degenerate quark mass $m_0$ and the isoscalar vector interaction reads~\cite{Buballa:2003qv}
\begin{equation}
\begin{split}
    \mathcal{L}_{\rm NJL} &=\bar{\psi}(i\gamma^\mu\partial_\mu-m_0)\psi\\
    &+ G_S[(\bar{\psi}\psi\,)^2+(\bar{\psi}i\gamma_5\vec{\tau}\psi\,)^2]-G_V(\bar{\psi}\gamma_\mu\psi)^2\textrm.
\end{split}
\end{equation}
In the mean-field approximation, the effective thermodynamic potential is given by~\cite{Buballa:2003qv}
\begin{equation}
\Omega_{\rm NJL}=\frac{(M-m_0)^2}{4G_S}-\frac{(\mu-\tilde{\mu})^2}{4G_V} + \Omega_{\rm vac}+\Omega_{\rm th}\textrm, \label{eq:pnjl_potential}
\end{equation}
where 
\begin{eqnarray}
   \Omega_{\rm vac}=-12\int\limits_{|\boldsymbol{p}|<\Lambda}\frac{\dd^3p}{(2\pi)^3} \eps_q
\end{eqnarray}
with $\eps_q=\sqrt{\boldsymbol{p}^2+M^2}$ is the vacuum contribution which we regularize with the sharp three-momentum cutoff. The thermal contribution reads
\begin{eqnarray}
   \Omega_{\rm th}=12T\int\frac{\dd^3p}{(2\pi)^3}\big( \ln(1-f) + \ln(1-\bar f)\,\big)
\end{eqnarray}
and does not require regularization since the integral is convergent. The Fermi-Dirac distributions for quark and antiquarks read
\begin{equation}
    f = \left(e^{(\eps_q-\tilde\mu)/T} + 1\right)^{-1} \textrm, \quad \bar f = \left(e^{(\eps_q+\tilde\mu)/T} + 1\right)^{-1}
\end{equation}
respectively. The gap equations can be obtained from the effective potential~\cite{Buballa:2003qv}, 
\begin{equation}
    \frac{\partial \Omega_{\rm NJL}}{\partial M}=0\textrm, \qquad \frac{\partial \Omega_{\rm NJL}}{\partial \tilde{\mu}}=0.
\end{equation}

In general, the phase diagram obtained in the NJL model, and thus the existence and location of the critical point is regularization scheme dependent~\cite{Kohyama:2015hix}. In this work, we use two different schemes, for which the parameters can be seen in Tab.~\ref{Tab:NJL_param}. The first scenario yields a CP at $\mu_B\approx990\,$MeV and $T\approx19\,$MeV
(Set 1), while the second does not predict the existence of
a CP (Set 2).

In Fig.~\ref{Fig:alpha_NJL_schemes}, we show the expansion parameter $\alpha_2$ obtained in these two models. Both parameters increase with temperature and exhibit very similar behavior, even though only one of the models predicts the CP. Consequently, from the lowest-order expansion, there is no sign of criticality from the CP, although one of the models (Set 1) features a CP in the phase diagram.

\begin{table}[t]
    \centering
    \begin{tabular}{c|c|c|c|c}
        set & $G_S\Lambda^2$ & $\Lambda~$[MeV]  & $m_0~$[MeV] & source   \\
       \hline\hline
        1 & 2.083 & 665.0 & 5.0 & \cite{Kohyama:2015hix} \\
        2 & 2.060 & 664.3 & 5.0 & \cite{Buballa:2003qv}\\
        3 & 4.3479 & 364.69 & 5.0 & this work
    \end{tabular}
    \caption{Parameters of the NJL model used in this work.}
    \label{Tab:NJL_param}
\end{table}

\begin{table*}[]
\centering
\begin{tabular}{c|c|c|c|c|c|c|c}
$\hat G_V$ &
  $T_0^c~${[}GeV{]} &
  $\alpha_2'(T^c_0)~${[}GeV$^{-2}${]} &
  $\alpha_2(T^c_0)~${[}GeV$^{-1}${]} &
  $\mu_B^{\rm pred.}~${[}GeV{]} &
  $T_c^{\rm pred.}~${[}GeV{]} &
  $\mu_B^{\rm NJL}~${[}GeV{]} &
  $T_c^{\rm NJL}~${[}GeV{]} \\\hline\hline
0.000 & 0.16097 & -74.398 & -0.1925 & 0.164     & 0.158 & 0.100 & 0.160 \\
0.075 & 0.16096 & -12.41  & -0.1316 & 0.401     & 0.150 & 0.330 & 0.153 \\
0.100 & 0.16094 & -5.279  & -0.1182 & 0.616     & 0.139 & 0.477 & 0.146 \\
0.125 & 0.16090 & -1.883  & -0.1066 & 1.031     & 0.104 & 0.618 & 0.137 \\
0.167 & 0.16040 & 0  & -0.091 & --     & -- & 0.789 & 0.124 \\
0.250 & -- & --   & --  & -- & --    & 1.008 & 0.104
\end{tabular}
\caption{Results obtained by applying the constant-entropy expansion to the NJL model (parameter set 3) for different values of the dimensionless repulsive coupling $\hat G_V$. Predicted critical baryon chemical potential and temperature are compared with the results obtained directly from the model.}
\label{tab:njl_gv_res}
\end{table*}

\begin{figure}
    \centering
    \includegraphics[width=1\linewidth]{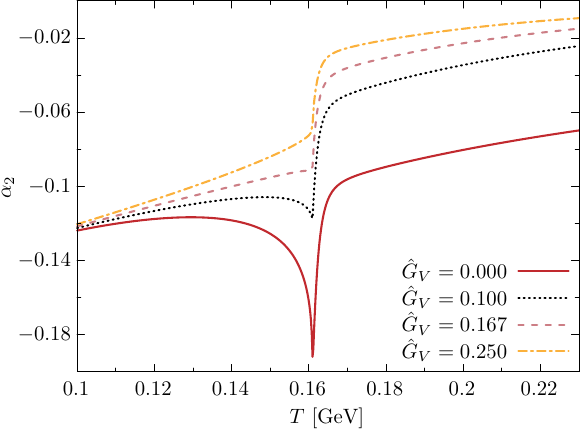}
    \caption{The expansion parameter $\alpha_2$ in the NJL model for different values of the dimensionless vector coupling $\hat G_V$.}
    \label{Fig:alpha_NJL}
\end{figure}

Another well-known feature of the NJL model is that the increase of the vector coupling moves the critical point towards lower temperatures until it vanishes for the critical value of the coupling~\cite{Buballa:2003qv, Sasaki:2006ws}. We use these features of the model to adjust the location of the critical point and study the predictive power of the expansion~\eqref{eq:taylor}. If the critical point lies close to $\mu_B=0$, where the expansion was performed, its effect is expected to appear even in the lower order coefficients. However, as the CP moves to larger chemical potentials, its effect on the $\mu_B=0$ thermodynamics is reduced, and one needs higher (if not infinite) orders in the expansion to be sensitive to the criticality~\cite{Karsch:2010hm}. This explains, why parameter sets 1 and 2 of the NJL model lead to almost the same $\alpha_2$.

To further study the expansion truncated at the second order, we consider an artificial parameterization (see the bottom row of Tab.~\ref{Tab:NJL_param}) of the NJL model chosen such that
the critical point is located at exceptionally low $\mu_B\approx100\,$MeV and high $T\approx160$\,MeV for vanishing vector coupling $G_V=0$. We stress that such parametrization is not expected to match the known phenomenology, e.g., the pion mass or the quark condensate, but serves merely as a tool to study the expansion under consideration. By increasing $\hat G_V \equiv G_V/G_S$ for a fixed value of $G_S$, the CP shifts towards smaller temperatures and higher chemical potentials and vanishes for the critical value of the coupling $\hat G_V=0.94$.

The $\alpha_2$ curves for several values of $\hat G_V$ are shown in Fig.~\ref{Fig:alpha_NJL}. We observe strong sensitivity to the value of the vector coupling. Particularly, the dip seen in the vicinity of the pseudo-critical temperature becomes shallower as $\hat G_V$ increases and vanishes for $\hat G_V=0.167$. 

 \begin{figure}
    \centering
    \includegraphics[width=1\linewidth]{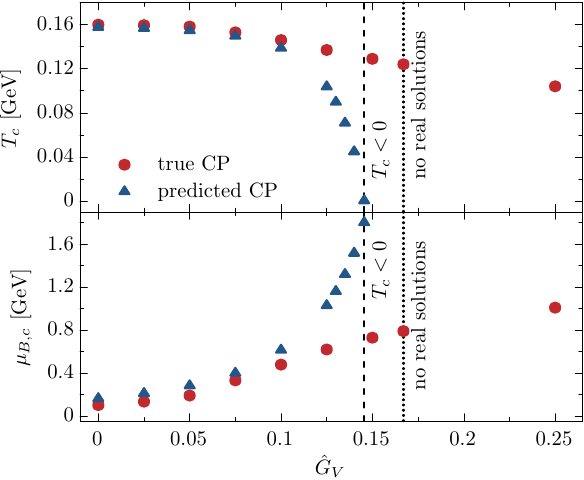}
    \caption{The location of the true (circles) and predicted (triangles) critical points in terms of temperature (top) and baryon chemical potential (bottom) as a function of the dimensionless repulsive coupling $\hat G_V$ in the NJL model. The vertical dashed line marks $\hat G_V$ above which the predicted $T_c$ is negative. The vertical dotted line marks $\hat G_V$ above which there is no prediction for the CP in the truncated NJL model (see text for details).}
    \label{Fig:NJL_CPs}
\end{figure}

By utilizing Eqs.~\eqref{eq:CP_criterion_alpha}, we obtained predictions for the CP for different values of $\hat G_V$. Our findings are summarized in Tab.~\ref{tab:njl_gv_res}. We observe that predictions for the location of the CP become worse as $\hat G_V$ increases, i.e. CP becomes more distant from the $\mu_B=0$ line. For $\hat G_V=0$ the prediction for $\mu_B$ differs by approximately $60\,$MeV, while for $\hat G_V=0.125$ the difference exceeds $400\,$MeV. The predictions for $T_c$ are more accurate but also worsen with the increasing vector coupling. Particularly, for $\hat G_V\gtrsim0.145$, the method predicts $T_c < 0$, and for $\hat G_V\gtrsim0.167$, the method fails to predict the existence of the CP. This behavior can be understood as follows: for small values of $\hat G_V$, there is an inflection point in $\alpha_2$ with $\dd\alpha_2/\dd T<0$, which gives a prediction for the CP at finite and positive $\mu_B$. As $\hat G_V$ increases, the dip becomes shallower and disappears, hence $\dd\alpha_2/\dd T\to 0^-$, and the predicted value of the baryon chemical potential approaches $\infty$, while the predicted value of $T\to -\infty$. In the present setup, this happens for $\hat G_V=0.167$. 
 
 For larger values of the vector coupling, we observe that $\alpha_2$ is monotonically increasing in temperature, and there is no prediction for a real baryon chemical potential, even though there is a CP in the full model. In Fig.~\ref{Fig:NJL_CPs} we compare the actual and predicted values of $T$ (upper panel) and $\mu_B$ (lower panel) for different values of $\hat G_V$. The vertical dashed line marks $\hat G_V$ above which the predicted $T_c$ is negative. The dashed line marks the value of the coupling above which Eqs.~\eqref{eq:CP_criterion_alpha} have no real solutions. If the CP is sufficiently far, the expansion cannot account for its effect in the lowest order. This highlights the limited validity of this approximation. We emphasize that our artificial parametrization demonstrates the limitation of the truncated expansion qualitatively, however, it does not provide a quantitative prediction for the range of its usability.
 
 We note that, due to the absence of the dynamical mesons in the NJL model, $\alpha_2$ behaves similarly to the HRG result with only the baryons present, tending to a negative value for $T\to0$. On the other hand, if the Polyakov loop is included for the effective confinement, one has $\alpha_2\to0$ as the temperature vanishes. Therefore, there is an additional inflection point, which always gives a prediction for the CP, even if the model has no criticality. However, this inflection point and the predicted CP have no connection to the chiral transition. 

 Finally, one can also include meson fluctuations in the quark-meson (QM) model -- which gives qualitatively the same results as the NJL model -- in an ideal-gas or Gaussian approximation~\cite{Kovacs:2021kas}. Similarly to the Polyakov loop, the dynamical mesons lead to $\alpha_2 \to 0$ for $T\to0$ in agreement with the HRG results. Consequently, the expansion in such an extended QM model predicts the existence of a CP for any parametrization of the model, although it is not related to the chiral phase transition.

\section{Conclusion}

We have considered the constant-entropy-density-contour expansion scheme, which was recently proposed for the search of the putative critical point in the QCD phase diagram~\cite{Shah:2024img}. We have applied the scheme to the hadron resonance gas model, which does not predict the existence of any CP because of the lack of criticality built into it. Nevertheless, we find that the expansion scheme truncated at the second-order Taylor coefficient yields a prediction for the CP at finite temperature and baryon chemical potential. By analyzing the expansion coefficient, we found that the prediction of the CP is due to the presence of both mesonic and baryonic degrees of freedom, and thus is not necessarily a signal of a CP.

We have also considered the Nambu--Jona-Lasinio model which captures the spontaneous breaking of the chiral symmetry and its restoration and thus may predict the existence of CP, depending on the parameterization. Utilizing different regularization schemes, we demonstrated that the expansion at the lowest truncation order is not sensitive to the critical effects unless the critical point is located exceptionally close to the $\mu_B=0$ line. We also found that the method under consideration may fail to predict the existence of the CP, even though it is present in the full model. This spotlights its limited applicability and suggests that further careful analysis of the expansion scheme is required.

\section*{Acknowledgments}
The authors are thankful for the valuable discussion with P.~M.~Lo, K.~Redlich, and C.~Sasaki. M.~M. also acknowledges helpful comments and discussion with J.~Goswami. The work of M.~M. was supported through the program Excellence Initiative–Research University of the University of Wroc\l{}aw of the Ministry of Education and Science. This work was supported by the Polish National Science Centre (NCN) under Preludium grant 2020/37/N/ST2/00367 (M.~S.) and OPUS Grant No. 2022/45/B/ST2/01527 (G.~K.).

\appendix

\bibliography{biblio}

\end{document}